\documentclass[amssymb,prd,superscriptaddress,aps,nofootinbib,twocolumn,showpacs]{revtex4-1}
\usepackage{graphicx}
\usepackage{amssymb,epsfig}
\usepackage{amsmath}
\usepackage[british]{babel}
\usepackage[breaklinks]{hyperref}
\usepackage{color}
\newcommand{\tn}{\textnormal}

\def\apj{Astrophysical Journal }                

\begin{document}
\title{Rotating black holes in Einstein-dilaton-Gauss-Bonnet gravity\\ with finite coupling}

\author{Andrea Maselli}
\affiliation{Center for Relativistic Astrophysics, School of Physics, Georgia Institute of Technology, Atlanta, GA
  30332, USA.}
\author{Paolo Pani}
\affiliation{Dipartimento di Fisica, Universit\`a di Roma ``La Sapienza'' \& Sezione INFN Roma1, P.A. Moro 5, 00185,
  Roma, Italy.}
\affiliation{CENTRA, Departamento de F\'{\i}sica, Instituto Superior T\'ecnico, Universidade de Lisboa, Avenida~Rovisco
  Pais 1, 1049 Lisboa, Portugal.}
\author{Leonardo Gualtieri}
\affiliation{Dipartimento di Fisica, Universit\`a di Roma ``La Sapienza'' \& Sezione INFN Roma1, P.A. Moro 5, 00185,
  Roma, Italy.}
\author{Valeria Ferrari }
\affiliation{Dipartimento di Fisica, Universit\`a di Roma ``La Sapienza'' \& Sezione INFN Roma1, P.A. Moro 5, 00185,
  Roma, Italy.}

\begin{abstract} 
Among various strong-curvature extensions of general relativity,
Einstein-dilaton-Gauss-Bonnet gravity stands out as the only nontrivial theory
containing quadratic curvature corrections while being free from the
Ostrogradsky instability to any order in the coupling parameter.  We derive an
approximate stationary and axisymmetric black hole solution of this
gravitational theory in closed form, which is of fifth order in the black hole
spin and of seventh order in the coupling parameter of the
theory.  This extends previous work that obtained the corrections to the metric
only to second order in the spin and at the leading order in the coupling
parameter, and allows us to consider values of the coupling parameter close to
the maximum permitted by theoretical constraints. We compute some quantities
which characterize this solution, such as the dilaton charge, the moment of inertia
and the quadrupole moment, and its geodesic structure, including the
innermost stable circular orbit and the epicyclic frequencies for massive
particles. The latter provides a valuable tool to test general relativity
against strong-curvature corrections through observations of the electromagnetic
spectrum of accreting black holes.  \end{abstract}

\maketitle

\section{Introduction}

Observational and experimental tests of general relativity (GR) 
\cite{Will:2014xja}
have mostly probed the weak-field/slow-motion regimes of the theory, while a
number of strong-field, relativistic GR predictions still remain elusive and difficult to
verify~\cite{Psaltis:2008bb,Yunes:2013dva}.
Furthermore, a series of long-lasting problems in Einstein's theory -- such as
the accelerated expansion of the Universe, dark matter, the nature of curvature
singularities, and the quest for an ultraviolet completion of GR -- have
motivated strong efforts to develop extended theories of gravity which would
modify GR in its most extreme regimes while conforming with current weak-field
observations~\cite{OleMiss}.

Black holes (BHs) are genuine strong-field predictions of GR and have no analog
in Newtonian theory. Thus, they are natural candidates to test gravity in the
strong-field regime.  Future networks of electromagnetic
detectors~\cite{2008SPIE.7013E..2AE,gendreau2012neutron,feroci2012large,2014ApJ...784....7B}
and ground-based gravitational-wave
detectors~\cite{2008CQGra..25k4045A,2010CQGra..27h4006H} will allow us to
measure some crucial properties of BHs, such as their shadows, the location of
the event horizon, and of the innermost stable circular orbit (ISCO). This
information will be instrumental to test the {\it Kerr hypothesis}, according to
which all stationary astrophysical BHs are uniquely described by the Kerr
family and are, thus, characterized by only two parameters: their mass and
angular momentum (see, e.g., Ref.~\cite{wiltshire2009kerr} and references therein).

In recent years, several modified theories of gravity have been proposed. They
can be divided in various categories, each one lifting some of the fundamental
principles (Lorentz invariance, weak and strong equivalence principles, massless
spin-2 mediators, etc.) upon which Einstein's theory is uniquely
built~\cite{OleMiss}. From this and other classifications, it emerges that one
of the simplest and best motivated ways to modify GR consists of including a
fundamental scalar field which is nonminimally coupled to the metric tensor. In
order to modify the strong-curvature regime, it is natural to couple this scalar
field in the gravitational action to terms quadratic in the curvature tensor.
Such couplings can also be interpreted as the first terms in the expansion in all
possible curvature invariants, as suggested by low-energy effective string
theories~\cite{Gross:1986mw}.  Generally, a quadratic curvature term in the
action leads to field equations of third (or higher) order, which are subject to
Ostrogradsky's instability~\cite{2007LNP...720..403W}. Therefore, these theories
should be considered as effective, i.e., truncations of a theory with further
terms in the action, which are neglected in the perturbative regime.

It should also be mentioned that quadratic curvature terms are crucial, not only
to modify the strong-curvature regime of GR, but also to affect the behavior of
stationary BHs. Indeed, standard scalar-tensor theories (in which one or more
scalar fields are included in the gravitational sector of the action) satisfy
the so-called no-hair theorems; i.e., stationary, vacuum BHs are the same as in
GR \cite{Bekenstein:1995un,heusler1996black,Sotiriou:2011dz} (but
see~\cite{Babichev:2013cya,Herdeiro:2014goa,Dias:2015rxy} for possible
violations of these theorems). When quadratic curvature terms are included in
the action instead, stationary BH solutions are different.

We shall consider a member of this family of modified-gravity theories, Einstein-dilaton-Gauss-Bonnet (EDGB) theory,
in which a scalar field (the dilaton) is coupled to the Gauss-Bonnet invariant~\cite{Kanti:1995vq,Gross:1986mw}
\begin{equation}
{\cal R}^2_{\tn{GB}}=R_{\alpha\beta\gamma\delta}R^{\alpha\beta\gamma\delta}
-4R_{\alpha\beta}R^{\alpha\beta}+R^2 \label{GB}
\end{equation}
in the action. EDGB gravity is one of the best motivated alternatives to GR.
Indeed, it is the only theory of gravity with quadratic curvature terms in the
action, whose field equations are of second differential order for {\em any}
coupling, and not just in the weak-coupling limit which is assumed in the
effective-field-theory approach~\cite{OleMiss}.  As a consequence, EDGB gravity
is ghost-free; i.e., it avoids the Ostrogradsky
instability~\cite{2007LNP...720..403W}. Furthermore, as mentioned above, the
higher-curvature coupling -- which modifies the strong-curvature regime of
gravity -- violates the hypothesis of the BH no-hair theorems so that BH
solutions in EDGB gravity are different from those predicted by GR and provide
the ideal arena for genuine strong-field tests of the Kerr hypothesis. Finally,
the EDGB term naturally arises in low-energy effective string
theories~\cite{Moura:2006pz}.

In this work, we construct an analytical, perturbative solution of EDGB theory,
which describes a slowly rotating BH endowed with a scalar field. To this aim,
we extend the formalism developed in~\cite{Hartle:1967he,Hartle:1968si} up to
fifth order in the BH (dimensionless) spin parameter $\chi=J/M^2$, where $J$ and
$M$ are the angular momentum and the Arnowitt-Deser-Misner mass of the solution,
respectively.

Analytical BH solutions of EDGB theory in the small-coupling limit have been
investigated in~\cite{Mignemi:1992nt,Yunes:2011we}, where stationary,
spherically symmetric configurations where found. Approximate, stationary, and axisymmetric
solutions to linear and quadratic order in the BH spin were obtained in
\cite{Pani:2011gy} and~\cite{Ayzenberg:2014aka}, respectively.  Both of these works
considered a weak-field expansion of the coupling between the scalar field and
the Gauss-Bonnet invariant ${\cal R}^2_{\tn{GB}}$ in terms of a dimensionful
coupling constant $\alpha$.
Exact numerical solutions were constructed to zeroth~\cite{Kanti:1995vq} and
first order~\cite{Pani:2009wy} in the spin  and also for arbitrary values of the
angular momentum~\cite{Kleihaus:2011tg,Kleihaus:2014lba}. Although exact in
$\alpha$, such solutions are of limited practical use (for instance, for Monte
Carlo data analysis) because they require a numerical integration for each set
of parameters. On the other hand, numerical solutions are necessary in regimes
where the slow-spin expansion does not converge and are, therefore, complementary
to our analysis.

Our results extend the study carried out so far. In particular, we go beyond the
analysis of Ref.~\cite{Ayzenberg:2014aka}, where  a BH solution was obtained  to
second order both in the spin and in the coupling parameter. Indeed, we compute
the metric tensor and the scalar field up to ${\cal
O}\left(\zeta^7,\chi^5\right)$, where $\zeta\equiv\alpha/M^2$, and $\alpha$ is
the EDGB coupling constant.  We use this expansion to derive the main
features  of the solution, such as the geometry of the event
horizon and  of  the ergoregion.  Furthermore, we study
the geodesic structure of this solution by computing the ISCO and the epicyclic
frequencies (see, e.g., Refs.~\cite{wald2010general,1999MNRAS.304..155M,Abramowicz:2002iu}) 
consistent with our approximation scheme. We compare these quantities with those obtained
in~\cite{Maselli:2014fca}, where a numerical solution was derived, which is
exact in the coupling parameter (i.e., with no perturbative expansion in
$\zeta$) and approximate to linear order in the BH spin.  We find relative
errors at most of the order of $1\%$ for the maximum value of $\zeta$ allowed by
theoretical constraints for the existence of BH solutions $\zeta\lesssim
0.691$~\cite{Kanti:1995vq} and much smaller for less extreme
couplings.

The results of this paper can be useful to devise tests of GR in the
strong-field regime through astrophysical observations of BHs. For instance, we
have shown~\cite{Maselli:2014fca} that observations of quasi-periodic
oscillations of accreting BHs, with the sensitivity of recently proposed
large-area X-ray space telescopes
(e.g.,~\cite{feroci2012large,gendreau2012neutron}), allow us to set 
constraints on the parameter space of EDGB theory, thus, probing the strong-field
regime of gravity (see, also, Ref.~\cite{Moore:2015bxa} for a recent study).
However, since BH solutions in EDGB theory (for finite $\alpha$) were only known
at first order\footnote{As mentioned above, a solution for finite spin and
coupling is only known in numerical
form~\cite{Kleihaus:2011tg,Kleihaus:2014lba}, and it is impractical for
extensive studies of geodesic properties. However, numerical
solutions are necessary to explore the high-spin regime, especially because EDGB
BHs can violate the Kerr bound and can have $\chi>1$~\cite{Kleihaus:2011tg}.}
in the spin parameter $\chi$, in~\cite{Maselli:2014fca} we only considered BHs
with very slow-rotation rate, for which the deviations from GR are expected to
be small.

This paper is organized as follows. In Sec.~\ref{Sec:EDGB_field} we derive
our solution of the EDGB field equations, describing rotating BHs up to ${\cal
O}\left(\zeta^7,\chi^5\right)$. In Sec.~\ref{sec:results} we study this
solution, computing its geometrical properties, the location of
the ISCO, and the azimuthal and epicyclic frequencies. We also estimate the accuracy
of our approximation in the determination of these quantities and how our
results improve on the existing literature. In particular, we discuss how the
spin correction to the azimuthal and epicyclic frequencies can affect possible
tests of GR based on observations of accreting BHs, such as those discussed
in~\cite{Maselli:2014fca}. Finally, in Sec.~\ref{concl} we draw our
conclusions.

\section{Spinning black holes in Einstein-Dilaton-Gauss-Bonnet theory}\label{Sec:EDGB_field}
In this section we derive the spacetime metric and scalar field describing rotating BHs in EDGB theory, up to
${\cal O}\left(\zeta^7,\chi^5\right)$.
\subsection{EDGB gravity}
Einstein-dilaton-Gauss-Bonnet theory is defined by the following action~\cite{Gross:1986mw,Kanti:1995vq}:
\begin{equation}\label{EDGB}
S=\frac{1}{2}\int
d^{4}x\sqrt{-g}\left[R-\frac{1}{2}\partial_{\mu}\Phi\partial^{\mu}\Phi
  +\frac{\alpha e^\Phi}{4}{\cal R}^2_{\tn{GB}}\right]\ ,
\end{equation}
where $g<0$ is the metric determinant, $\Phi$ is a scalar field coupled to the Gauss-Bonnet invariant~\eqref{GB} and
$\alpha>0$ is the coupling constant~\cite{Kanti:1995vq}.  Since we are interested in BH solutions, in the action above
we have neglected matter fields.
We use geometric units $G=c=1$: with this choice, the scalar field $\Phi$ is dimensionless and $\alpha$ has the
dimensions of a length squared.

The field equations of EDGB gravity are found
by varying the action~\eqref{EDGB} with respect to $g_{\mu\nu}$ and $\Phi$:
\begin{eqnarray}
&&G{^{\mu}}_{\nu}=\frac{1}{2}\partial^{\mu}\Phi\partial_{\nu}\Phi-
\frac{1}{4}g{^\mu}_{\nu}\partial_{\alpha}\Phi\partial^{\alpha}\Phi-\alpha{\cal K}{^{\mu}}_{\nu}\,,\label{EinsEq}\\
&&{\cal S}\equiv \frac{1}{\sqrt{-g}}\partial_{\mu}(\sqrt{-g}\partial^{\mu}\Phi)+\frac{\alpha}{4}e^{\Phi}
{\cal R}^2_{\tn{GB}}=0\,, \label{KGeq}
\end{eqnarray}
where $G_{\mu\nu}=R_{\mu\nu}-\frac{1}{2}g_{\mu\nu}R$ is the Einstein tensor,
\begin{eqnarray}
{\cal K}_{\mu\nu}&=&\frac{1}{8}\left(g_{\mu\rho}g_{\nu\lambda}+g_{\mu\lambda}g_{\nu\rho}\right)\epsilon^{k\lambda\alpha\beta}\nonumber\\
&&\times\nabla_\gamma\left(\epsilon^{\rho\gamma\mu\nu}R_{\mu\nu\alpha\beta} \partial_{k}e^\Phi\right)\,,
\end{eqnarray}
and $\epsilon^{\mu\nu\alpha\beta}$ is the Levi-Civita tensor, with
$\epsilon^{0123}=-(-g)^{-1/2}$. Note that -- by virtue of the GB combination
entering the action~\eqref{EDGB} -- the equations are of second differential
order, and, therefore, this theory is free from the Ostrogradsky
instability~\cite{2007LNP...720..403W}. Indeed, EDGB gravity is a particular
case~\cite{Kobayashi:2011nu} of Horndeski gravity -- the most general
scalar-tensor theory with second-order field equations~\cite{Horndeski:1974wa}.
This special subcase is the only one known to date in which regular, stationary,
asymptotically flat, hairy BH solutions other than GR ones are
found~\cite{Sotiriou:2013qea}. Furthermore, EDGB gravity can be obtained from
the low-energy expansion of the bosonic sector of heterotic string
theory~\cite{Gross:1986mw,Moura:2006pz}, in such case the coupling $\alpha$ is
related to the string tension.

In order to simplify our notation, in the next sections we shall introduce the
modified Einstein tensor $\tilde{G}{^{\mu}}_\nu=G{^{\mu}}_\nu-T{^\mu}_\nu$,
where
\begin{equation} T{^\mu}_\nu=\frac{1}{2}\partial^{\mu}\Phi\partial_{\nu}\Phi
-\frac{1}{4}g{^\mu}_{\nu}\partial_{\alpha}\Phi\partial^{\alpha}\Phi-\alpha{\cal
K}{^{\mu}}_{\nu}\,, \end{equation}
is the effective stress-energy tensor for the dilaton.

\subsection{Static BH solutions}\label{Sec:Background}
Since the EDGB coupling constant has the dimensions of the inverse of the curvature tensor, it is natural to expect that
in this theory the strongest deviations from GR will come from physical systems involving high curvature, such as BHs,
neutron stars and the early Universe. We focus here on BH solutions and, in particular, on rotating BH geometries that
are obtained through a slow-rotation expansion around a static background solution.

The exact BH background solution (first derived in~\cite{Kanti:1995vq}) is described by the static, 
spherically symmetric line element
\begin{equation}\label{Schwarz}
ds^2=-e^{\Gamma(r)}dt^2+e^{-\Lambda(r)}dr^2
+r^2d\Omega^2\,,
\end{equation}
and by a spherically symmetric scalar field, $\Phi=\phi(r)$.
The field equations (\ref{EinsEq}) and (\ref{KGeq}) supplied by the metric 
ansatz (\ref{Schwarz}) reduce to a set of differential equations for the 
scalar field and for the functions $\Gamma$ and $\Lambda$. Indeed, 
Eq.~(\ref{KGeq}) yields
\begin{align}
  \phi''+\phi'\bigg(&\frac{\Gamma'-\Lambda'}{2}+\frac{2}{r}\bigg)=
  \frac{\alpha e^\phi}{2r^2}\bigg(\Gamma'\Lambda'e^{-\Lambda}\ +\nonumber\\
&+(1-e^{-\Lambda})\left[\Gamma''+\frac{\Gamma'}{2}(\Gamma'-\Lambda')\right]\bigg)\ ,
\end{align}
while the $t$-$t$, $r$-$r$ and $\theta$-$\theta$ components of $\tilde
G{^\mu}_{\nu}=0$ 
reduce to
\begin{align}\label{Gtt}
\bigg[1+\frac{\alpha e^\phi}{2r}\phi'(1-3e^{-\Lambda})\bigg]\Lambda'&=\frac{\phi'^2r}{4}
+\frac{1-e^\Lambda}{r}\ +\nonumber\\
+\frac{\alpha e^\phi}{r}(1-&e^{-\Lambda})(\phi''+\phi'^2)\ ,\\
\bigg[1+\frac{\alpha e^{\phi}}{2r}\phi'(1-3e^{-\Lambda})\bigg]\Gamma'&
=\frac{\phi'^2r}{4}+\frac{e^{\Lambda}-1}{r}\ ,\label{Grr}\\
\Gamma''+\left(\frac{\Gamma'}{2}+\frac{1}{r}\right)(\Gamma'-\Lambda')&
=-\frac{\phi'^2}{2}\ +\frac{\alpha e^{\phi-\Lambda}}{r}\ \times\nonumber\\
\times\bigg[\phi'\Gamma''+\Gamma'(\phi''+\phi'^2)&+\frac{\Gamma'\phi'}{2}
  (\phi'-3\Lambda')\bigg]\label{Gthth}\ .
\end{align}
Note that Eqs.~\eqref{Gtt}--\eqref{Gthth} are not all independent and that the $r$-$r$ component can be solved
analytically, yielding
\begin{equation}
e^{\Lambda}=\frac{-\beta+\sqrt{\beta^2-4\gamma}}{2}\ ,
\end{equation}
where
\begin{equation}
\beta=\frac{\phi'^2 r^2}{4}-1-\Gamma'\left(r+\frac{e^\phi\phi'}{2}\right)\,,\quad \gamma=\frac{3}{2}\Gamma'\phi'e^{\phi}\,.
\end{equation}
The remaining two independent equations can be written as 
\begin{equation}\label{eqmetric}
\phi''=-\frac{d_{1}}{d}\quad \ ,\quad \ \Gamma''=-\frac{d_{2}}{d}\ ,
\end{equation}
where the radial functions $d,d_{1}$ and $d_{2}$ are given in Appendix A of~\cite{Kanti:1995vq}.  The
Arnowitt-Deser-Misner mass $M$ and the dilatonic charge ${\cal D}$ can be read off the asymptotic behavior of the metric
and of the dilaton field,
\begin{eqnarray}
 g_{tt}&\to& -1+2M/r+\dots \\
 \phi&\to& {\cal D}/r +\dots 
\end{eqnarray}
It turns out that for each value of $M$, there is only one solution describing a static BH. In other words, the scalar
field is a ``secondary hair'': the dilatonic charge ${\cal D}$ is not an independent parameter but is determined in
terms of the BH mass $M$.

The field equations are invariant under the rescaling $\phi\rightarrow\phi+\hat\phi$ and $r\rightarrow re^{\hat\phi/2}$
(or, equivalently, $M\rightarrow Me^{\hat\phi/2}$ and ${\cal D}\rightarrow {\cal D}e^{\hat\phi/2}$) where $\hat\phi$ is a
constant. We fix this freedom by requiring that $\phi\rightarrow 0$ at spatial infinity; this means that at infinity,
the Gauss-Bonnet invariant appears in the action~\eqref{EDGB} multiplied by the constant $\alpha/4$.

As noted in~\cite{Kanti:1995vq} static BH solutions in EDGB gravity exist only if 
\begin{equation}\label{alphabound1}
e^{\phi_{\tn{h}}}\le\frac{r_{\tn{h}}}{\alpha\sqrt{6}}\ ,
\end{equation}
where $\phi_\tn{h}$ is the value of the scalar field computed at the horizon $r_{\tn{h}}$. As shown
in~\cite{Pani:2009wy}, by requiring that $\phi\rightarrow 0$ at spatial infinity, Eq.~(\ref{alphabound1}) can be recast
in the form
\begin{equation}
0\leq\frac{\alpha}{M^2}\lesssim 0.691\,. \label{rangealpha}
\end{equation}
Thus, smaller BHs would correspond to a more stringent bound on $\alpha$. 

Presently, the tightest observational bound on the EDGB coupling parameter (obtained by the orbital decay of X-ray
binaries) is $\alpha\lesssim 47M_\odot^2$~\cite{Yagi:2012gp}. As discussed in \cite{Maselli:2014fca}, this upper bound
is weaker than the theoretical constraint~\eqref{alphabound1} for BHs with $M\lesssim8.2M_\odot$. For such BHs, the
entire range~\eqref{rangealpha} is phenomenologically allowed.

Solutions of Eqs.~\eqref{eqmetric} have been solved numerically in Ref.~\cite{Kanti:1995vq}, while an analytical static
BH solution has been derived to second order in $\alpha/M^2$ in Refs.~\cite{Mignemi:1992nt,Yunes:2011we}.

\subsection{Spinning BH solutions}\label{Sec:Rotating}
To describe slowly rotating BH solutions we extend the approach developed by Hartle~\cite{Hartle:1967he,Hartle:1968si},
in which spin corrections to the static solutions are introduced within a perturbative framework. The procedure described in this section is generic and can be applied also to other theories and different spinning solutions.

Let us start with the most general solution for a stationary, axially symmetric spacetime\footnote{We also assume
  equatorial symmetry and invariance under $(t\to -t,\varphi\to-\varphi)$.} which is given by
\begin{align}\label{Hartle1}
ds^2=-H^2dt^2+Q^2&dr^2+r^2K^2[d\theta^2\ + \nonumber\\
&+\sin^2\theta(d\varphi-Ldt)^2]\ ,
\end{align}
where $H,Q,K,$ and $L$ are functions of $(r,\theta)$. The ansatz~\eqref{Hartle1} can be expanded perturbatively in the spin
around the static solution
\begin{eqnarray}\label{Hartle2}
ds^2=&&-e^{\Gamma}[1+2h(r,\theta)]dt^2+e^{-\Lambda}[1+2m(r,\theta)]dr^2 \nonumber\\
&&+r^2[1+2k(r,\theta)][d\theta^2+\sin^2\theta(d\varphi-\hat\omega(r,\theta)dt)^2]\,,\nonumber\\
\end{eqnarray}
where the functions $\hat\omega$, $h$, $m$, and $k$ can be expanded 
in a complete basis of orthogonal functions according to their symmetry properties as 
\begin{align}
\hat\omega&=\sum_{n=1,3,5,...}^{N_\chi-q} \sum_{l=1,3,5,...}^n \chi^n \omega_{l}^{(n)}(r) S_l(\theta) \,, \label{omegahat}\\
h&=\sum_{n=2,4,...}^{N_\chi-p} \sum_{l=0,2,4,...}^n \chi^n h_{l}^{(n)}(r) P_l(\cos\theta)\ ,\\
m&=\sum_{n=2,4,...}^{N_\chi-p} \sum_{l=0,2,4,...}^n \chi^n m_{l}^{(n)}(r) P_l(\cos\theta)\ ,\\
k&=\sum_{n=2,4,...}^{N_\chi-p} \sum_{l=0,2,4,...}^n \chi^n k_{l}^{(n)}(r) P_l(\cos\theta)\ ,
\end{align}
where $P_l$ are the Legendre polynomials, 
$S_l=-\frac{1}{\sin\theta}\frac{dP_{l}(\cos\theta)}{d\theta}$ (note that
$P_0=S_1=1$), and $p$ (respectively $q$) is zero when the order $N_\chi$ of the spin
expansion is even (respectively odd), whereas $p=1$ (resp. $q=1$) otherwise.  The radial
functions $\left(\omega_{l}^{(n)},h_{l}^{(n)},m_{l}^{(n)},k_{l}^{(n)}\right)$
are of the order $\mathcal{O}(\chi^n)$.  Note that, since the
metric~\eqref{Hartle2} is invariant under the rescaling $r\rightarrow f(r)$, the
functions $k_0^{(n)}(r)$ can be set to zero without loss of
generality~\cite{Hartle:1967he,Hartle:1968si}.
 
Because the dilaton field transforms as a scalar under rotation, we expand it as
\begin{equation}
\Phi(r)=\phi(r)+\sum_{n=2,4,...}^{N_\chi-p} \sum_{l=0,2,4,...}^n \chi^n
\phi_{l}^{(n)}(r) P_l(\cos\theta) \ , 
\end{equation} where $\phi$ is the
background static solution and the radial functions $\phi_{l}^{(n)}$ are of the
order $\mathcal{O}(\chi^n)$.


\subsubsection{$\mathcal{O}(\chi)$ corrections}

Rotating BH solutions in EDGB gravity have been investigated to linear order in the spin angular momentum in Refs.~\cite{Pani:2009wy,Pani:2011gy}. 
At this order, the metric (\ref{Hartle2}) reduces to the static case with a
nonvanishing gravitomagnetic term described by $\hat\omega(r,\theta)=\omega_1^{(1)}\equiv \omega(r)$ [see Eq.~\eqref{omegahat}]:
\begin{eqnarray}
ds^2&=&-e^{\Gamma(r)}dt^2+e^{-\Lambda(r)}dr^2+r^2d\Omega^2\nonumber\\
&&-2r^2\omega(r)\sin^2\theta dtd\varphi\,.
\end{eqnarray}
From $\tilde{G}^t_{~\varphi}=0$, it is easy to show that $\omega$ satisfies the second-order equation~\cite{Pani:2009wy}:
\begin{eqnarray}\label{gtphieq}
\omega''&&+\left[2r^2e^\Lambda-2\alpha r\phi'e^\phi\right]^{-1}\nonumber\\
&&\left(-\alpha e^\phi\left[2\phi''r+\phi'(6+2\phi'r-\Gamma'r-3\Lambda'r)\right] \right.\nonumber\\
&&\left.-e^\Lambda r\left[-8+r(\Gamma'+\Lambda')\right]\right)\omega'=0\,,
\end{eqnarray}
where the coefficient of $\omega'$ depends on the nonspinning solution. The BH
angular momentum can be read off the asymptotic behavior of the gyromagnetic
term, \begin{equation}\label{boundary} \omega(r)\to \frac{2J}{r^3}\,,
\end{equation} at large distance.
\subsubsection{$\mathcal{O}(\chi^{n})$ corrections: ${n}\geq2$ and
even}\label{Sec:Second_order}
Replacing the metric ansatz~\eqref{Hartle2} into the field equations and using
the decomposition in Legendre polynomials,  a set of ordinary
differential equations can be obtained, at each order in the spin expansion. The equations are
inhomogeneous with source terms given by the lower-order functions. 
%

At each given order ${n}\geq2$ with even ${n}$, the equations are found
 from $E_1\equiv\tilde{G}_{tt}=0$, $E_2\equiv\tilde{G}_{rr}=0$,
$E_3\equiv\tilde{G}_{\theta\theta}+(\sin\theta)^{-2}
\tilde{G}_{\varphi\varphi}=0$, and $E_4\equiv{\cal S}$ for the scalar
equation~(\ref{KGeq}), each  contracted with a Legendre
polynomial,
\begin{equation} \int_0^\pi d\theta \sin\theta P_l(\cos\theta)
E_i(r,\theta)=0\,, \end{equation}
where $i=1,2,3,4$ and $l=0,2,4,...,{n}$. Because of the symmetry properties of the
field equations and of the background, this procedure gives a set of purely
radial, inhomogeneous, ordinary differential equations for $h_{l}^{({n})}$,
$m_{l}^{({n})}$, $k_{l}^{({n})}$, and $\phi_l^{({n})}$ with $l=0,2,4,\dots,{n}$ (we
recall that $k_{0}^{({n})}=0$).  

%

\subsubsection{$\mathcal{O}(\chi^{n})$ corrections: ${n}\geq3$ and odd}
Similarly, at a given order ${n}\geq3$ (with odd ${n}$) in the spin expansion, a
set of radial equations for the gravitomagnetic terms can be obtained by
contracting $\tilde{G}_{t\varphi}=0$ with the (axisymmetric) vector spherical
harmonics, namely
\begin{equation} \int_0^\pi d\theta \sin\theta
\frac{dP_l(\cos\theta)}{d\cos\theta} \tilde{G}_{t\varphi}=0 \end{equation}
with $l=1,3,5,...,{n}$. Again, this procedure yields a set of purely radial,
inhomogeneous, ordinary differential equations for $\omega_{l}^{({n})}$ with
$l=1,3,5,\ldots,{n}$.  


\subsection{Small-coupling approximation}\label{Sec:small_coupling}

The set of equations presented above provides a full description of the BH
solution  at any perturbative  order in the spin,  but generic
(i.e., nonperturbative) in the EDGB coupling. However, such equations are
cumbersome and it is impractical to solve them numerically. More importantly,
the theoretical constraint~\eqref{rangealpha} shows that the dimensionless
coupling parameter has to be smaller than unity. This motivates a small-coupling
approximation~\cite{Pani:2009wy,Pani:2011gy,Ayzenberg:2014aka}, in which the
field equations are solved perturbatively in $\alpha/M^2\ll1$ to some desired
order. Actually, because we are interested in the regime $\alpha/M^2\lesssim1$
(the maximum value\footnote{Note that the constraint
$\zeta\equiv\alpha/M^2\lesssim0.691$ is valid for nonspinning solutions at
finite coupling. The precise value of the upper bound can be modified for large
rotation rates~\cite{Kleihaus:2011tg}. Indeed, as discussed later in this
section, the BH mass acquires ${\cal O}(\chi^2)$ corrections which can be
reabsorbed in the definition of the mass. Nonetheless, the bound on
$\alpha/M^2$  emerges only from the nonperturbative BH  solutions
and does not appear in the small-coupling approximation (to any order in $\zeta$).} of this
parameter is $0.691$), we shall compute terms of relatively high order in this
expansion.

To simplify the notation, we introduce the dimensionless parameter
\begin{equation}
\zeta=\frac{\alpha}{M^2}\,.
\end{equation}
As a result of our approximation scheme, we expand all quantities, such as the metric functions and the scalar field, in
terms of the two parameters $\zeta\,,\,\chi$. For example,
\begin{equation}
g_{\mu\nu}=g_{\mu\nu}^{(0,0)}+\sum_{i=1}^{N_\chi}
\sum_{j=1}^{N_\zeta}\chi^i \zeta^jg_{\mu\nu}^{(i,j)}\ ,
\end{equation}
where the double superscript $^{(i,j)}$ denotes the order of the expansion in
the BH spin parameter and in the EDGB coupling parameter, respectively;
$g_{\mu\nu}^{(0,0)}$ is the Schwarzschild metric.  In practice, using the spin
decomposition previously discussed, we simply expand the set of radial variables
$\vec{f}=\{\Gamma,\Lambda,\phi,\omega_{l}^{(n)},m_{l}^{(n)},h_{l}^{(n)},k_{l}^{(n)},\phi_{l}^{(n)}\}$
as \begin{equation} f=\sum_{j=0}^{{N_\zeta}}\zeta^{j}f^{(j)}\ , \end{equation}
where $f^{(j)}$ are radial functions which do not depend on the coupling
parameter $\zeta$.  By replacing these expressions into the field equations
derived in Sec.~\ref{Sec:Rotating}, and solving them order by order in $\zeta$,
we obtain the desired expansion for the metric tensor and the scalar field.
Remarkably, this procedure yields an analytical solution. We compute the
explicit solution up to $\mathcal{O}(\zeta^7,\chi^5)$, but the
procedure can be straightforwardly extended to higher order both in $\zeta$ and
in $\chi$.

Solving the differential equations at each order in $\chi$ and $\zeta$ yields some
integration constants, which are uniquely fixed by requiring that
\begin{enumerate}
\item the metric is asymptotically flat, and the scalar field vanishes at spatial infinity;
\item there exists an event horizon, where perturbations are regular;
\item the {\it physical} mass and angular momentum of the BH are given by $M$ and $M^2\chi$, as measured by an observer
  at spatial infinity. In particular, the bare mass of the $\mathcal{O}(\zeta^0)$ solution acquires some corrections to
  each order in $\zeta$, which are reabsorbed in the physical mass $M$.
\end{enumerate}
We note that only one of the two integration constants appearing in the solution of the scalar field at each order in $\zeta$ is fixed by requiring regularity outside the horizon, while the metric is regular for each value of the remaining constants. Although this is not evident in the Schwarzschild coordinates adopted here, it can be nonetheless checked by computing some curvature invariants. However, the remaining integration constants can all be reabsorbed in the definitions of the physical mass and angular momentum, so that the final solution truncated at a given order depends only on two parameters, as in the Kerr case.

The explicit expressions of the metric tensor and the scalar field up to $\mathcal{O}(\zeta^7,\chi^5)$ are quite
long and are available in a {\scshape Mathematica} notebook provided in the
Supplemental Material. For completeness, the explicit Kerr metric to ${\cal O}(\chi^5)$ in the Hartle-Thorne coordinates is given in Appendix~\ref{app:Kerr5}.

\section{Geometrical and geodesic properties of the solution}\label{sec:results}
Here we study the properties of the analytical solution we have derived. To this aim, we compute some geometrical and
geodesic quantities which characterize the spinning EDGB BH solution to $\mathcal{O}(\zeta^7,\chi^5)$.

\subsection{Event horizon, ergosphere, intrinsic curvature, and dilaton charge}

The event horizon is given by the largest root $r=r_{\tn{h}}$ of the equation (cf. e.g.,~\cite{poisson2004relativist})
$g_{\phi\phi} g_{tt}-g_{t\phi}^2=0$, which yields the following power expansion in terms of $\zeta$ and $\chi$:
\begin{align}\label{Eqrh}
\frac{r_{\tn{h}}}{M}=\sum_{i=0}^7\zeta^i(a_{i}+b_{i}\chi+c_{i}\chi^2+d_{i}\chi^3+e_{i}\chi^4+f_{i}\chi^5)\ ,
\end{align}
where the coefficients $(a_{i},b_{i},c_{i},d_{i},e_{i},f_{i})$ are listed\footnote{For the sake of clarity, the coefficients shown in the Appendix will be rounded to some numerical
  factors. The exact expressions are available in the Supplemental Material {\scshape Mathematica}
  notebook.} in Table~\ref{TableCoeff} of Appendix~\ref{app:coeff2}.
As in the Kerr case, the horizon radius $r_{\tn{h}}$ does not depend on the angular
coordinates. Nonetheless, its intrinsic geometry -- as computed by considering a
spatial section $dt=0$ at $r=r_{\tn{h}}$ -- is nonspherical. Indeed,
\begin{equation} ds^2_{t={\rm const},
r=r_{\tn{h}}}=g_{\theta\theta}(r=r_{\tn{h}},\theta)d\Omega^2\,, \label{line2D}
\end{equation}
and since $g_{\theta\theta}$  explicitly depends on
$\theta$, the intrinsic geometry is nonspherical. For the line
element~\eqref{line2D}, the curvature radius is
\begin{eqnarray} R_{\rm intr} &=& \frac{2}{g_{\theta\theta}}-\frac{\cot\theta
g_{\theta\theta}'}{g_{\theta\theta}^2}+\frac{g_{\theta\theta}'^2}{g_{\theta\theta}^3}
-\frac{g_{\theta\theta}''}{g_{\theta\theta}^2} \,, \end{eqnarray}
where (only in the above formula) a prime denotes differentiation with respect to
$\theta$, and  for our solution is
\begin{align}\label{curvrad} M^2
R_\tn{intr}=&\sum_{i=0}^7\zeta^i[l_{i}+\chi^2(m_{i}+n_{i}\cos2\theta)\
+\nonumber\\ &+\chi^4(p_{i}+q_{i}\cos2\theta+u_{i}\cos4\theta)]\ ; \end{align}
this is constant only when $\zeta=0=\chi$. Hereafter, we adopt  the same
expansion of Eqs.~(\ref{Eqrh}) and \eqref{curvrad} for other physical
quantities. The numerical values of the coefficients of these expansions are given in
Appendix~\ref{app:coeff2}, whereas their exact form is provided in the
Supplemental Material {\scshape Mathematica} notebook.

The location of the ergosphere is given by the largest root of $g_{tt}=0$,
\begin{align}\label{EqERGO}
\frac{r_\tn{ergo}}{M}=&\sum_{i=0}^7\zeta^i[l_{i}+\chi^2(m_{i}+n_{i}\cos2\theta)\ +\nonumber\\
&+\chi^4(p_{i}+q_{i}\cos2\theta+u_{i}\cos4\theta)]\,,
\end{align}
where the only nonvanishing spin corrections correspond to even powers of $\chi$.


Finally, the dilaton charge ${\cal D}$ can be extracted from the leading-order large-distance 
behavior of the dilaton field $\Phi\to {\cal D}/r$ and reads
\begin{align}
\frac{{\cal D}}{M}=&\sum_{i=1}^7\zeta^i(a_{i}+b_{i}\chi+c_{i}\chi^2+e_{i}\chi^4)\ ,
\label{charge}
\end{align}
where the coefficients $d_i$ and $f_i$ identically vanish.

\subsection{Moment of inertia}
The moment of inertia is defined as $I=J/\Omega_\tn{h}$, where $J$ is the BH angular momentum and $\Omega_\tn{h}$ is the angular velocity at the horizon of locally nonrotating observers,
\begin{equation}
 \Omega_\tn{h}=-\lim_{r\to r_\tn{h}}\frac{g_{t\varphi}}{g_{\varphi\varphi}}\,.
\end{equation}
In our case we obtain
\begin{widetext}
\begin{align}
  \frac{I}{M^3}=&\phantom{+}4-0.2625000\zeta ^2-0.1721966\zeta ^3 -0.1458764\zeta ^4-0.1409996 \zeta ^5
  -0.1474998 \zeta ^6-0.1627298 \zeta ^7\nonumber\\
 &-\chi^2[1 - 0.2359276 \zeta^2 - 0.2175544 \zeta^3 - 0.2431079 \zeta^4 - 0.2776072 \zeta^5 - 0.3283860 \zeta^6 \ +\nonumber\\
 &-0.3984877 \zeta^7]+\chi^4[0.25 - 0.1170266 \zeta^2 - 0.04956483 \zeta^3 +  0.01732049 \zeta^4 + 0.09842336 \zeta^5
\ + \nonumber\\
&+0.2055222 \zeta^6 +  0.3503737 \zeta^7]  \ , \label{Ibar}
\end{align}
\end{widetext}
where again the only nonvanishing spin corrections correspond to even powers of $\chi$.

\subsection{Quadrupole moment}
According to the BH no-hair theorems, the quadrupole moment (as well as the
higher-order multipole moments) of any regular, stationary, asymptotically flat
BH in GR is uniquely determined by its mass $M$ and angular momentum
$J$~\cite{Carter71,Hawking:1973uf,Hansen:1974zz}. A deformed Kerr geometry as
the one just discussed, does not necessarily possess this unique no-hair
property. Since the dilaton charge of this solution is not an independent
parameter, the multipole moments of an EDGB BH can all be written in terms of
$M$ and $J$, but the relations among them will change with respect to Kerr.  The
$\zeta$ corrections to the BH quadrupole moment are, thus, relevant to test the
Kerr hypothesis~\cite{Psaltis:2008bb,Yunes:2013dva,OleMiss}.

To compute the quadrupole moment, we follow the general approach described
in~\cite{1980RvMP...52..299T}, in which the multipole moments of an
asymptotically flat geometry are read off the asymptotic behavior of the metric.
This approach requires the metric to be expressed in asymptotically Cartesian
and mass-centered (ACMC) coordinates.  In particular, in order to extract the
quadrupole moment, the metric has to be ACMC-2, i.e., $g_{tt}$ and $g_{ij}$
($i,j\neq t$) should not contain any angular dependence up to
$\mathcal{O}(1/r^2)$ terms. In our case, the coordinate transformation that
enforces such property is 
\begin{eqnarray} r &\to& r+\frac{\chi^2M^2}{2 r} \left[1+\frac{M}{r}-\frac{2
M^2}{r^2}+\frac{M(6 M-r)}{r^2} \cos ^2\theta\right]\,, \nonumber\\ \theta &\to&
\theta +\frac{\chi^2 M^3}{r^3} \sin\theta \cos\theta\,, \nonumber \end{eqnarray}
and does not involve the EDGB coupling $\zeta$ or spin corrections higher than
second order. In the new ACMC-2 coordinates, the $g_{tt}$ component reads
\begin{align} g_{tt}=-1+\frac{2M}{r}&+\frac{\sqrt{3}}{2r^3}[Q_{20}Y^{20}\
+\nonumber\\ &+(l=0\ \tn{pole})]+\mathcal{O}\left(\frac{M^4}{r^4}\right)\ ,
\end{align} 
where $Y_{20}$ is the $(l=2,m=0)$ spherical harmonic, and $Q_{20}$ is
the $m=0$ mass quadrupole moment. From our explicit solution, we obtain to order
$\mathcal{O}(\zeta^7,\chi^5)$,
\begin{widetext} 
\begin{align} 
Q_{20}=&-\sqrt{\frac{64 \pi}{15}}\chi^2 M^3
\left[ \left(1+0.1061619\zeta^2+0.07524246\zeta^3
+0.07459416\zeta^4+0.07756926\zeta^5\right.\right.\ +\nonumber\\
&\left.\left.+0.08553316\zeta^6+0.09805643\zeta^7\right)-  \chi^2\zeta^2\left(0.0308519+0.0408857\zeta 
+0.0638894\zeta^2\ +\right.\right.\nonumber\\ 
&\left.\left.+0.0866408\zeta^3+0.116314\zeta^4+0.154763\zeta^5\right)\right]\,.
\label{Qbar} 
\end{align}
 \end{widetext}
Interestingly, the ${\cal O}(\chi^4)$ corrections to the quadrupole moment are
proportional to $\zeta^2$; i.e., they vanish in the GR limit. For $\zeta\sim0.4$,
the $\mathcal{O}(\zeta^2,\chi^2)$ correction to the quadrupole moment relative
to the Kerr case is about $1.7\%$, whereas the $\mathcal{O}(\zeta^3,\chi^2)$
correction is approximately $0.5\%$. Finally, for $\zeta\sim0.4$ and
$\chi\sim0.6$, the $\mathcal{O}(\zeta^2,\chi^4)$ correction is approximately
$0.1\%$.

We remark that the quadrupole moment of spinning
EDGB BHs has been computed numerically in \cite{Kleihaus:2014lba}. 
Our solution has the advantage of giving this quantity in analytical
form.


\subsection{Geodesics and epicyclic frequencies}

We shall now consider timelike geodesics in the slowly rotating EDGB BH
spacetime.  We assume a minimally coupled test particle and restrict to
equatorial orbits, for which $\theta=\pi/2$ and $d\theta=0$. We first compute
stable circular orbits; then, by considering small perturbations of these orbits, 
we derive the epicyclic frequencies $\omega_r$ and $\omega_\theta$ (see,
e.g., Refs.~\cite{wald2010general,1999MNRAS.304..155M,Abramowicz:2002iu}). For a
stationary-axisymmetric spacetime, the ISCO corresponds to the radius
at which the second derivative of the effective potential \begin{equation}
V(r)=\frac{1}{g_{rr}}\left(\frac{\mathcal{E}^2g_{\varphi\varphi}+2\mathcal{E}L
g_{t\varphi} +L^2g_{tt}}{g_{t\varphi}^2-g_{tt}g_{\varphi\varphi}}-1\right)
\end{equation}
vanishes. Here, we have introduced the particle-specific energy and angular
momentum $\mathcal{E}$ and $L$~\cite{1972ApJ...178..347B}, given by
\begin{align}
\mathcal{E}=&-\frac{g_{tt}+g_{t\varphi}\omega_\varphi}{\sqrt{-g_{tt}-2g_{t\varphi}\omega_\varphi-g_{\varphi\varphi}\omega_\varphi^2}}\
,\\
L=&\frac{g_{t\varphi}+g_{\varphi\varphi}\omega_\varphi}{\sqrt{-g_{tt}-2g_{t\varphi}\omega_\varphi-g_{\varphi\varphi}\omega_\varphi^2}}\
, \end{align} where $\omega_\varphi$ is the azimuthal angular velocity
\begin{equation}
\omega_\varphi=\frac{-g_{t\varphi,r}+\sqrt{g_{t\varphi,r}^2-g_{tt,r}g_{\varphi\varphi,r}}}{g_{\varphi\varphi,r}}\
.  \end{equation}
Solving $V''(r)=0$ order by order, we obtain the ISCO radius up to
$O(\zeta^7,\chi^5)$: \begin{align}\label{Isco}
\frac{r_\tn{ISCO}}{M}=\sum_{i=0}^7\zeta^i(a_{i}+b_{i}\chi&+c_{i}\chi^2+d_{i}\chi^3\
+\nonumber\\ &+e_{i}\chi^4+f_{i}\chi^5)\ .  \end{align}
Orbits with radius $r>r_\tn{ISCO}$ are stable. 
Under a small
perturbation,  a massive particle orbiting  in one of these stable, circular
orbits oscillates with radial and vertical frequencies given
by~\cite{wald2010general,1999MNRAS.304..155M,Abramowicz:2002iu}
\begin{align}\label{epi} 
\omega^2_r=&\frac{(g_{tt}+\omega_\varphi
g_{t\phi})^2}{2g_{rr}} \frac{\partial^2 {\cal U}}{\partial r^2}\bigg\vert_{l} \
, \\ \omega^2_\theta=&\frac{(g_{tt}+\omega_\varphi
g_{t\phi})^2}{2g_{\theta\theta}} \frac{\partial^2 {\cal U}}{\partial
\theta^2}\bigg\vert_{l}\ .  \end{align} 
These are the epicyclic frequencies.
Here ${\cal U}=g^{tt}-2l g^{t\varphi}+l^2g^{\varphi\varphi}$, with
$l=L/{\cal E}$ being the ratio between the particle angular momentum and its
energy~\cite{Abramowicz:2002iu}. The full expressions for
$\omega_r,\omega_\theta$, as well as for $\omega_\varphi$ as functions of
$(r,M,\chi,\zeta)$ and up to order $\mathcal{O}(\zeta^7,\chi^5)$ are
available in the {\scshape Mathematica} notebook
provided as Supplemental Material.  We explicitly show here their values at the
ISCO: \begin{align}
M\omega_{\varphi}\big\vert_\tn{ISCO}=\sum_{i=0}^7\zeta^i(a_{i}&+b_{i}\chi+c_{i}\chi^2+d_{i}\chi^3\
+\nonumber\\ &+e_{i}\chi^4+f_{i}\chi^5)\ ,\label{omegaphi}\\
M\omega_{\theta}\big\vert_\tn{ISCO}=\sum_{i=0}^7\zeta^i(a_{i}&+b_{i}\chi+c_{i}\chi^2+d_{i}\chi^3\
+\nonumber\\ &+e_{i}\chi^4+f_{i}\chi^5)\ , \label{omegatheta} \end{align}
whereas $\omega_{r}\big\vert_\tn{ISCO}=0$ as in the Kerr case.


\subsection{Comparison with previous results}\label{Sec:Comparison}

As a check, we can compare our results with those derived in~\cite{Ayzenberg:2014aka}, where the metric of the EDGB
spinning BH was found to ${\cal O}(\zeta^2,\chi^2)$ in Boyer-Lindquist coordinates. A direct comparison of the metric
coefficients is not possible, since the BH solutions have been derived on different charts. However, we can overcome
this problem by computing the Kretschmann invariant
$\mathcal{K}=R_{\alpha\beta\gamma\delta}R^{\alpha\beta\gamma\delta}$ and evaluating it at a specific point. From our
solution truncated at ${\cal O}(\zeta^2,\chi^2)$, we get
\begin{widetext}
\begin{align}\label{kret_scalar}
\mathcal{K}(r,\theta)=&\phantom{+}48\,{\frac{M^{2}}{r^6}}
+\frac{144M^2}{r^8}\left[(1-8\cos^2\theta)+\frac{M}{r}\sin^2\theta
+2\frac{M^2}{r^2}(3\cos^2\theta-1)\right]
-\frac{\zeta^2}{r^4}\left[\frac {2M^3}{r^3}+\frac{M^4}{r^4}+144\frac {M^5}{{r}^{5}}\ +\right.\nonumber\\
&\left.+14\frac {{M}^{6}}{{r}^{6}}+{\frac {128}{5}}{\frac {{M}^{7}}{{r}^{7}}}-1680
{\frac {{M}^{8}}{{r}^{8}}} \right]+\frac{\zeta^{2}\chi^{2}}{r^4}\left[{\frac {M^3}{{r}^{3}}}+{\frac {54431}{1750}}\,
{\frac {M^4}{{r}^{4}}}+{\frac {12846}{175}}\,{\frac {M^5}{{r}^{5}}}+{\frac {77047}{1225}}\frac{M^6}{r^6}
\ +\nonumber\right.\\
&\left.-{\frac {348909}{350}}\,{\frac {M^7}{{r}^{7}}}-{\frac 
{304938}{175}}\,{\frac {{M}^{8}}{{r}^{8}}}-{\frac {28023}{35}}\,{\frac {{M}^{9}}{{r}^{9}}}+{\frac {359468}{35}}{\frac {{M}^{10}}{{r}^{10}}}+{\frac {53848}{5}}
{\frac {{M}^{11}}{{r}^{11}}}-21984\,{\frac {{M}^{12}}{{r}^{12}}}\ +\nonumber\right.\\
&\left.+\left(-{\frac {80334}{875}}\,
{\frac {M^4}{{r}^{4}}}-{\frac {19638}{175}
}\,{\frac {M^5}{{r}^{5}}}-{\frac {234816}{1225}}\frac{M^6}{r^6}+{\frac {
1448877}{350}}\,{\frac {M^7}{{r}^{7}}}+{\frac {711114}{175}}\,{\frac {{
M}^{8}}{{r}^{8}}}+{\frac {92679}{35}}\,{\frac {{M}^{9}}{{r}^{9}}}\ 
+\right.\right. \nonumber\\
&\left.\left.-{
\frac {2168052}{35}}\,{\frac {{M}^{10}}{{r}^{10}}}-{\frac {59544}{5}}\,
{\frac {{M}^{11}}{{r}^{11}}}+65952\,{\frac {{M}^{12}}{{r}^{12}}}\right)\cos^2\theta\right]\ .
\end{align}
\end{widetext}
Replacing the explicit expression for $r_\textnormal{h}$ in Eq.~(\ref{Eqrh}), we find that on the horizon
\begin{align}
  \mathcal{K}(r_{\tn{h}},{\pi}/{2})=&\phantom{+}\frac{3}{4M^4}+\frac{9\chi^2}{8M^4}\ +\nonumber\\
  &+\frac{\zeta^2}{M^4}\left[\frac{327}{1280}+\frac{404023\chi^2}{784000}\right]\ .
\end{align}
This result coincides with the Kretschmann scalar derived in \cite{Ayzenberg:2014aka} and evaluated at the event
horizon on the equatorial plane in Boyer-Lindquist coordinates.  Finally, we have verified the agreement between the
expression for $M\omega_{\varphi}$ at the ISCO -- which is also a gauge invariant quantity -- obtained from the metric
derived in Ref.~\cite{Ayzenberg:2014aka}, and the same expression obtained truncating the expression in
Eq.~(\ref{omegaphi}) to $\mathcal{O}(\zeta^2,\chi^2)$.


\subsection{Accuracy of the expansion}\label{subsec:accuracy}

In this section, we estimate the accuracy of our perturbative scheme. In
particular, we estimate the truncation error arising from neglecting
$\mathcal{O}(\zeta^8)$ terms in the expansion. To this aim, we compare our
results with those obtained in Refs.~\cite{Kanti:1995vq,Pani:2009wy,Maselli:2014fca}, where a
solution for slowly rotating BHs in the EDGB theory has been derived at first
order in $\chi$ and is ``exact'' in $\zeta$ (i.e., with no perturbative
expansion in $\zeta$). To be consistent, we neglect terms of the order
$\mathcal{O}(\chi^2)$ in Eqs.~({\ref{Eqrh}}), (\ref{Isco}), (\ref{omegaphi})
and (\ref{omegatheta}).

\begin{figure}[ht]
\begin{center}
\includegraphics[width=7.5cm]{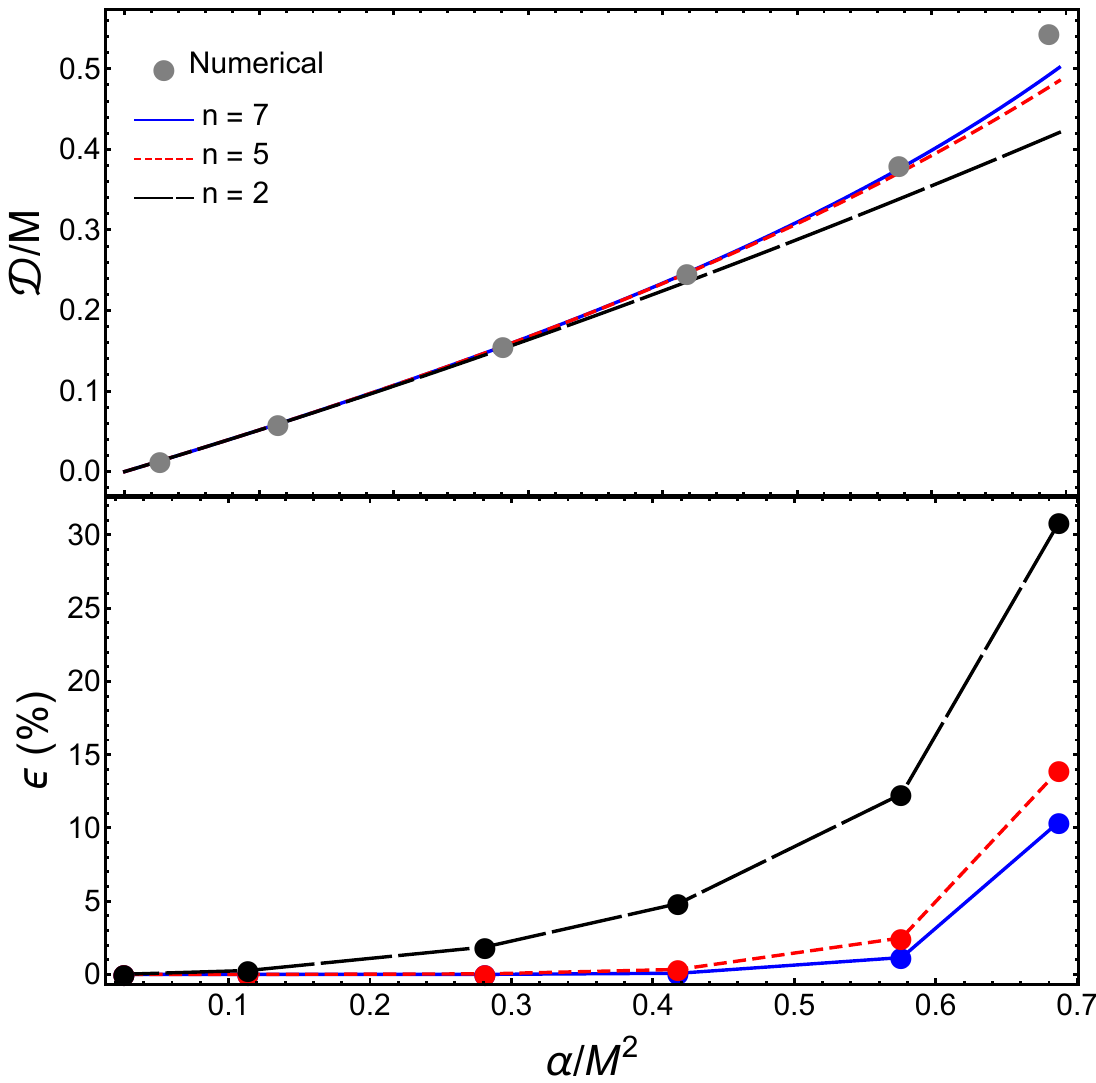}
\caption{Top panel: Dilatonic charge as a function of $\zeta=\alpha/M^2$ computed in~\cite{Kanti:1995vq,Pani:2009wy}
  (gray markers) compared with the expression in Eq.~\eqref{charge} truncated at ${\cal O}(\zeta^2)$, ${\cal
    O}(\zeta^5)$, and ${\cal O}(\zeta^7)$ and for vanishing spin. Bottom panel: Relative discrepancy between the
  perturbative and nonperturbative estimates of the dilatonic charge, as a function of $\alpha$ for various
  truncations.}
\label{fig:charge}
\end{center}
\end{figure}

In Fig.~\ref{fig:charge}, we compare the dilatonic charge computed
in~\cite{Kanti:1995vq,Pani:2009wy} nonperturbatively  in $\zeta$,
with the expression in Eq.~\eqref{charge}, for $\chi=0$, truncated at various
orders of $\zeta$. As expected, for $\zeta\ll 0.2$, higher-order corrections are
negligible, but they contribute significantly as $\zeta\to0.691$. To ${\cal
O}(\zeta^7)$, the deviation from the exact result is about $1\%$ for
$\zeta\sim 0.6$ and is as large as $5\%$ for $\zeta\sim 0.691$. In contrast, the
${\cal O}(\zeta^2)$ truncation differs by about $30\%$ as $\zeta$ increases
to its maximum value.

Likewise, for the set of quantities $f=\{r_\textnormal{h},r_\tn{ISCO},M\omega_\varphi\vert_\tn{ISCO},
M\omega_\theta\vert_\tn{ISCO}\}$, we compute the relative error
\begin{equation}
\epsilon_n=\frac{f^{(n,1)}}{\bar{f}}-1\,,\label{epsilon}
\end{equation}
where $\bar{f}$ represents the exact quantity (nonperturbative in 
$\zeta$)~\cite{Pani:2009wy,Maselli:2014fca}. We estimate $\epsilon_n$ at various orders of approximation in $\zeta$, for
different values of the BH spin parameter. 
In Table~\ref{Tableerrors}, we show the largest relative errors obtained for all considered quantities, at different levels of accuracy, in the
limiting case $\zeta=0.691$ (left) and $\zeta=0.576$ (right). We remark that
$\zeta=0.691$ is an extreme
situation, since for slightly smaller values of $\zeta$ (i.e., $\zeta=0.576$) the deviations are much smaller.

Figure~\ref{fig:charge} and Table~\ref{Tableerrors} show that our analytical
solution approaches the exact solution of \cite{Kanti:1995vq,Pani:2009wy} as
the value of $n$ increases, i.e., when we consider more and more terms in the
small-coupling expansion. In particular, for
$r_\tn{ISCO},M\omega_\varphi\vert_\tn{ISCO},M\omega_\theta\vert_\tn{ISCO}$, the
relative errors (for $n=7$) are always smaller than $1\%$ for any value of $\zeta$, even for
the maximum allowed value, $\zeta\sim0.691$.  For the horizon, the threshold
above which $\epsilon_{n=7}> 0.01$ is lower, namely $\zeta\sim
0.55 $.

\begin{table*}[t]
\centering
\begin{tabular}{c|cccccc}
\hline
\hline
&$\chi$ &  $\epsilon_{n=2}(\%)$ & $\epsilon_{n=4}(\%)$ & $\epsilon_{n=6}(\%)$  
& $\epsilon_{n=7}(\%)$  \\
\hline
$r_{\tn{h}}/M$& 0 & 5.90 & 4.45  & 3.72  & 3.48  \\
\hline
$r_\tn{ISCO}/M$& 0 & 1.00 & 0.58  & 0.43  & 0.39  \\
 &0.05 & 1.11 & 0.65  & 0.49  & 0.44  \\
 &0.10  & 1.23 & 0.72  & 0.54 & 0.49  \\
 \hline
$M\omega_\varphi\vert_\tn{ISCO}$& 0 & 1.36 & 0.79  & 0.59  & 0.53  \\
 &0.05 & 1.56 & 0.95  & 0.72  & 0.66  \\
 &0.10 & 1.88  & 1.22& 0.98  & 0.91  \\
 \hline
$M\omega_\theta\vert_\tn{ISCO}$& 0.05 & 1.53 & 0.92  & 0.69  & 0.63  \\
& 0.10 & 1.78 & 1.13  & 0.88  & 0.81  \\\hline
\hline
\end{tabular}
\begin{tabular}{c|cccccc}
\hline
\hline
&$\chi$ &  $\epsilon_{n=2}(\%)$ & $\epsilon_{n=4}(\%)$ & $\epsilon_{n=6}(\%)$  
& $\epsilon_{n=7}(\%)$  \\
\hline
$r_{\tn{h}}/M$& 0 & 1.33 & 0.52 & 0.24 & 0.17    \\
\hline
$r_\tn{ISCO}/M$& 0 & 0.32 & 0.093 & 0.038 & 0.026    \\
& 0.05 & 0.37 & 0.12 & 0.055 & 0.042    \\
& 0.1 & 0.42 & 0.14 & 0.074 & 0.059    \\
\hline
$M\omega_\varphi\vert_\tn{ISCO}$& 0 & 0.44 & 0.13 & 0.053 & 0.036    \\
& 0.05 & 0.56 & 0.23 & 0.14 & 0.13    \\
& 0.1 & 0.80 & 0.44 & 0.35 & 0.33    \\
\hline
$M\omega_\theta\vert_\tn{ISCO}$&  0.05 & 0.54 & 0.20 & 0.12 & 0.10    \\
& 0.1 & 0.71 & 0.36 & 0.27 & 0.25    \\
\hline
\hline
\end{tabular}
\caption{Left: the relative error $\epsilon_n$ [cf. Eq.~\eqref{epsilon}] between different quantities listed in the
  first column, computed through the solution derived in \cite{Pani:2009wy}, nonperturbative in $\zeta$, and compared
  with our perturbative results truncated at ${\cal O}(\zeta^n)$. We consider the maximum value of $\zeta$ allowed for
  BH solutions in EDGB gravity, $\zeta=0.691$, and different values of the BH spin parameter.
Right: Same  for $\zeta=0.576$.
}
\label{Tableerrors}
\end{table*}

\subsection{Are spin corrections important?}
The analysis presented in the Sec.~\ref{subsec:accuracy} shows that the
metric expanded in powers of $\zeta$, which we derived in a closed, analytic
form, is a very good approximation of the exact numerical result: it
reproduces the most relevant geodesic quantities within $1\%$ for the maximum
value $\zeta\sim0.691$ and within $0.3\%$ for $\zeta\sim0.576$. It is, therefore,
justified to adopt such higher-order perturbative expansion as a starting point
to devise strong-field tests of gravity.

\begin{figure*}
\begin{center}
\includegraphics[width=7.7cm]{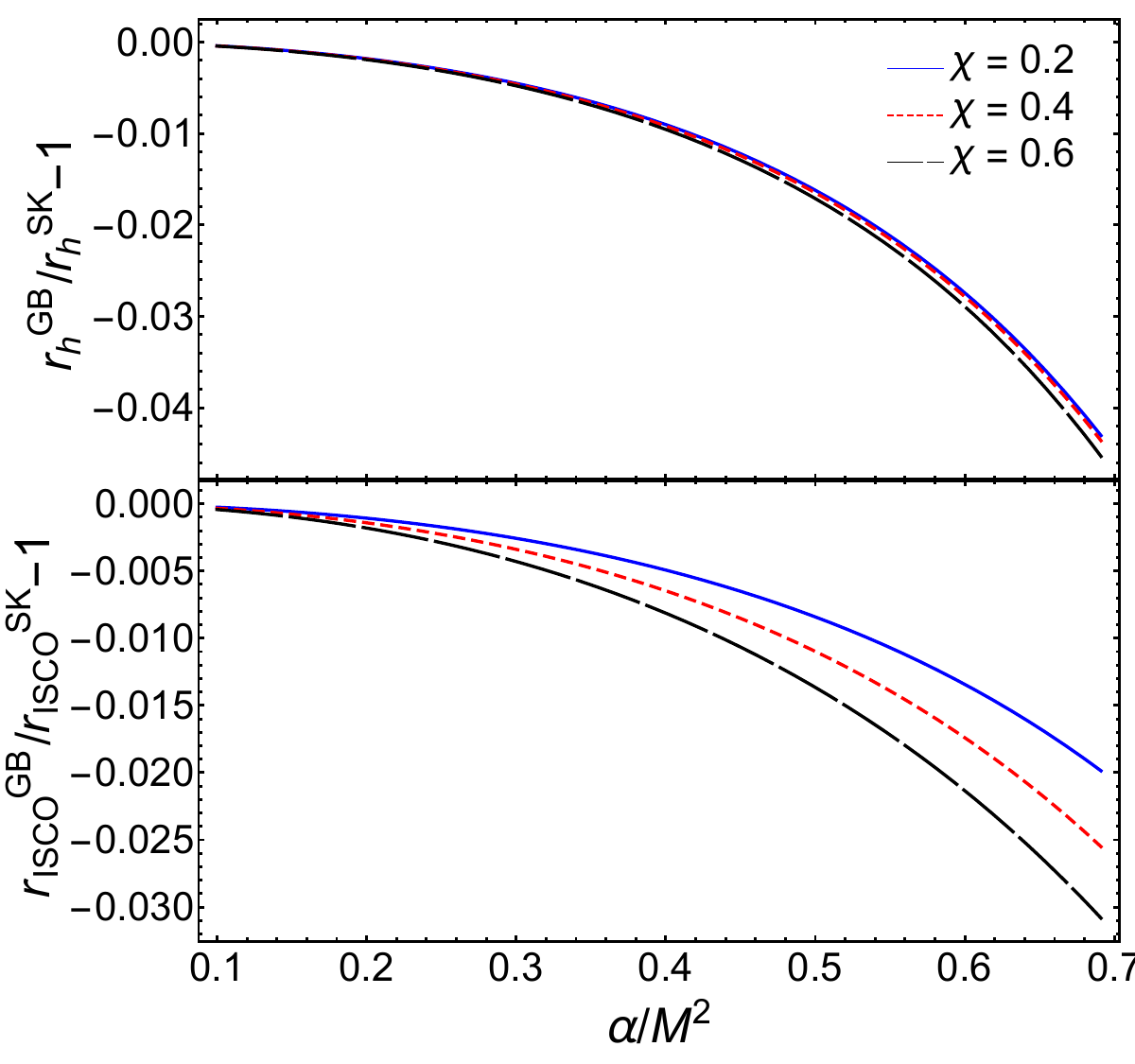}
\includegraphics[width=7.7cm]{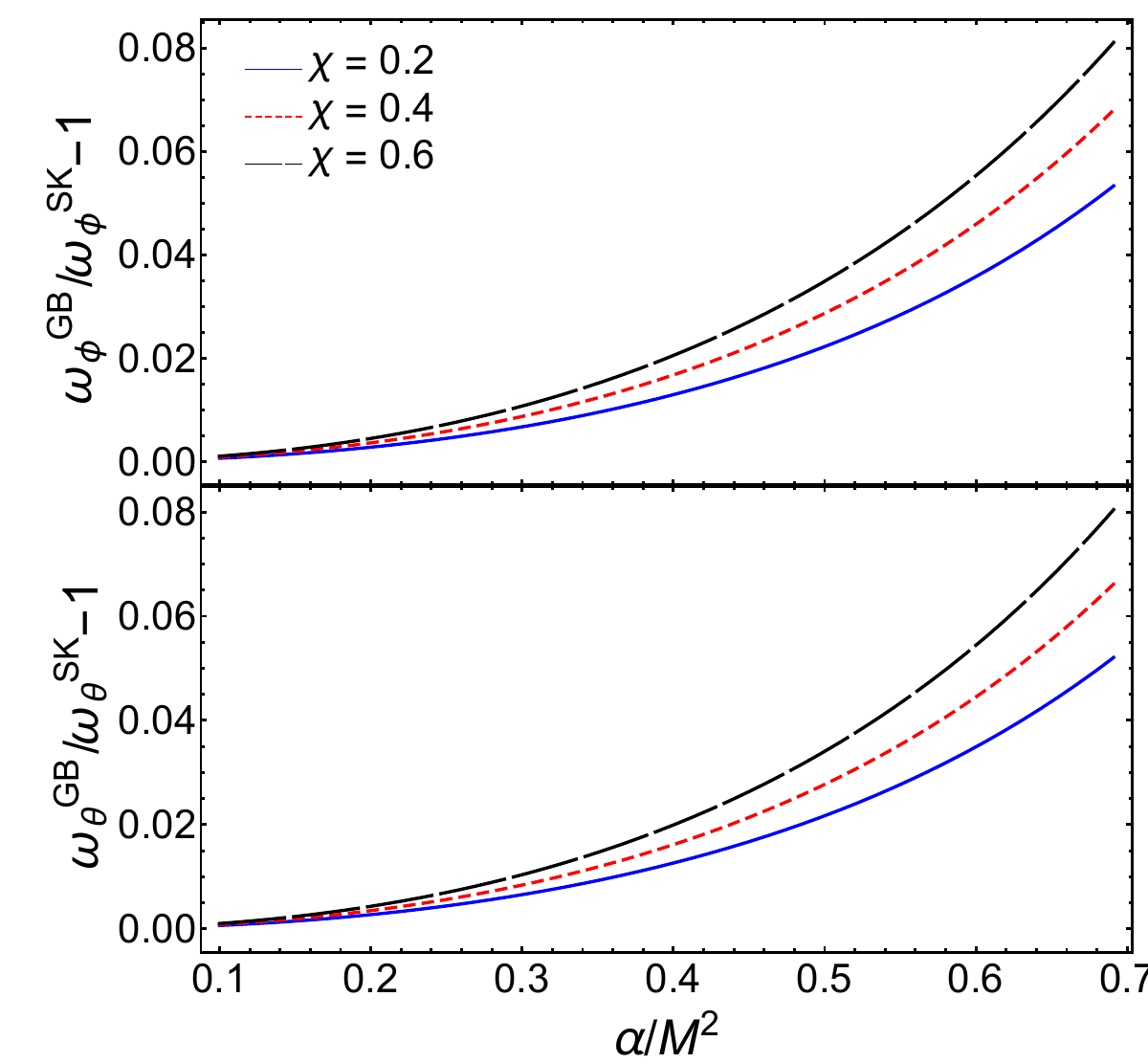}
\caption{Left panel: The percentual error in the horizon radius $r_{\tn{h}}$ and in the ISCO
$r_\tn{ISCO}$ for our perturbative result [Eqs.~\eqref{Eqrh} and~\eqref{Isco}],
relative to the Kerr solution expanded at ${\cal O}(\chi^5)$, as a function of
the EDGB coupling parameter $\alpha$. We consider three values of the BH
spin parameter $\chi=(0.2,0.4,0.6)$. Right panel: Same as the left panel but
for the epicyclic frequencies Eqs.~\eqref{omegaphi} and \eqref{omegatheta}
evaluated at the ISCO.}
\label{figHorIsco}
\end{center}
\end{figure*}

In Ref.~\cite{Maselli:2014fca}, we studied the deviations of the azimuthal and
epicyclic frequencies in a slowly rotating EDGB BH to first order in the spin.
However,  deviations from the Kerr case should
increase with higher values of the spin.  Indeed, as the spin 
increases, the ISCO gets closer to the horizon, and, therefore, observables from  orbits near the
ISCO probe a region of higher curvature, where the deviations should be larger.

In Fig.~\ref{figHorIsco}, we confirm this claim by showing the deviations of the
horizon and ISCO locations and, most important, of the azimuthal frequency
$\omega_\varphi$ and angular epicyclic frequency $\omega_\theta$ at the ISCO,
relative to their values computed using the Kerr metric approximated at ${\cal
O}(\chi^5)$, and as functions of $\zeta$ and $\chi$. For a fixed value of
$\zeta$, the percentual errors are systematically larger as the spin increases,
reaching up to $7\%$ for $\chi=0.6$. This large value of the spin parameter
should be considered as an extrapolation. Indeed, our results neglect terms of
the order ${\cal O}(\chi^6)$, which introduce corrections of roughly $5\%$ for
$\chi\sim0.6$.

Our perturbative solution is also useful to estimate the convergence
properties of the expansion. From the coefficients listed in
Table~\ref{TableCoeff}, we can compute the ratio of the ${\cal O}(\chi^n)$
and  ${\cal O}(\chi^{n-1})$ corrections for a given
quantity. For the angular epicyclic frequency $M\omega_{\theta}\vert_\tn{ISCO}$,
this ratio is roughly $(0.41,0.39,0.37)$ for $n=(3,4,5)$, in the
extreme case $\chi\sim0.5$ and $\zeta\sim 0.5$. Therefore, the fifth-order spin
correction is about 20\% of the quadratic  one. Other quantities show
a similar behavior. Clearly, the convergence improves for smaller values of
$\chi$, whereas it is almost insensitive to the values of the EDGB coupling
$\zeta$.

Finally, we note that the percentual error of the horizon location is almost
insensitive to the spin, whereas the epicyclic frequencies are much more
sensitive to this parameter.

\section{Concluding remarks}\label{concl}
With the advent of precision measurements of the spectrum of accreting compact
objects, it is of utmost importance to devise tests of gravity that use these
measurements to probe the geometry near compact objects.  To this aim, we have
considered a specific modified theory -- namely, EDGB gravity -- as a case study.
This theory has some appealing theoretical features; for example, it is free from
instabilities and circumvents the BH no-hair theorems. Furthermore, it modifies
GR precisely in the strong-curvature regions, while passing all current
solar-system and binary-pulsar tests~\cite{OleMiss}.

Spinning BHs in this theory have been studied in the past, both
numerically~\cite{Kanti:1995vq,Pani:2009wy,Kleihaus:2011tg,Kleihaus:2014lba} and
analytically~\cite{Mignemi:1992nt,Yunes:2011we,Pani:2011gy,Ayzenberg:2014aka} to
leading order in the coupling parameter. Numerical solutions have the advantage
of being general, but they are impractical for some applications, for example
for Monte Carlo simulations spanning a high-dimensional parameter space.
Approximate analytical solutions can be very useful for this purpose, although
they are usually perturbative.

Here, as a first step to develop precision tests of gravity based on geodesic
motion near stationary BHs, we have constructed an analytical, approximate
solution of EDGB theory describing a deformed Kerr BH. The solution is valid to
fifth order in the spin and to seventh order in the coupling parameter, thus,
extending previous solutions that are valid only to quadratic order in the
coupling and the spin. With the analytical solution at hand, it is
straightforward to compute various quantities of interest. We have presented the
corrections to the horizon and ergoregion location, moment of inertia and
quadrupole moment relative to the Kerr metric, as well as the charge of the
dilaton field that characterizes this solution.
For a given value of the coupling $\zeta$, the solution depends only on the mass
$M$ and on the dimensionless angular momentum $\chi$, while the dilaton charge
is fixed in terms of $M$.  In addition, we have computed some geodesic
quantities, namely, the ISCO location and the azimuthal and epicyclic frequencies
as functions of $M$, $\chi$, $\zeta$, and the orbital radius $r$.

When truncated at first order in the BH spin,
our solution reproduces the most
relevant geodesic quantities obtained in~\cite{Kanti:1995vq,Pani:2009wy}
with a numerical approach, within $1\%$ for the maximum value
$\zeta\sim0.691$,  and within $0.3\%$ for 
 $\zeta\sim0.576$. The accuracy of the solution grows
dramatically for smaller values of the coupling. These results indicate that our
perturbative solution is a good approximation of the exact numerical results.

In a future publication, we will extend the analysis of
Ref.~\cite{Maselli:2014fca}, which studied how observations of quasiperiodic
oscillations in the spectrum of accreting BHs can be used to constrain EDGB
theory in the strong-field regime, to larger values of the BH spin.  A
similar analysis can also be performed for other tests based on stationary
BHs, for example, those based on the broadened iron line (e.g.,
~\cite{Bambi:2012at,Jiang:2015dla}) or the continuum fitting method
(e.g.,~\cite{Bambi:2011jq}; see also,~\cite{Moore:2015bxa}).  On the technical
side, our perturbative approach is generic: it can be applied to any order in
$\zeta$ and in $\chi$, as well as to other modified-gravity theories and
different spinning solutions.

\begin{acknowledgments}
%
A.M. is supported by the NSF Grants No. 1205864, No. 1212433, and No. 1333360. 
P.P. was supported by the European Community through the Intra-European 
Marie Curie Contract No.~AstroGRAphy-2013-623439 and by FCT-Portugal 
through the Project No. IF/00293/2013.
This work was partially supported by 
the NRHEP Grant No. 295189 FP7-PEOPLE-2011-IRSES. 
\end{acknowledgments}
\newpage
\appendix

\section{Kerr metric in the Hartle-Thorne approximation}\label{app:Kerr5}
In this appendix, we show the form of the BH solution in the Hartle-Thorne approximation for $\alpha=0$, i.e., the slowly rotating Kerr BH in GR, up to the fifth order in the BH spin angular momentum,
\begin{widetext}
\begin{align}
g_{tt}=&\phantom{+}1-\frac{2 M}{r}+J^2 \left[\left(\frac{2}{M r^3}-\frac{2}{r^4}-\frac{4 M}{r^5}\right) P_2(\cos\theta)+\frac{2 \cos2\theta}{r^4}\right]+J^4\bigg\{\frac{2}{5 M^2r^6}-\frac{12 M}{5 r^9}+\frac{11}{5M r^7}\nonumber\\
&+\frac{6}{5 r^8}+\left[\frac{146}{7 r^8}-\frac{16}{7 M^2 r^6}+\frac{44 M}{7 r^9}+\frac{46}{7 M r^7}-\cos (2\theta ) \left(\frac{4}{M r^7}+\frac{8}{r^8}\right)\right]P_2(\cos\theta)\nonumber\\
&-\left(\frac{8}{5 M r^7}+\frac{24}{5 r^8}\right)\cos (2 \theta )+\sin ^2(\theta ) \left(\frac{8}{3 M^2 r^6}-\frac{8}{15 M r^7}-\frac{48}{5 r^8}\right)   S_3(\theta )\nonumber\\
&+\left(\frac{66}{35 M^2 r^6}-\frac{2}{M^3 r^5}-\frac{192 M}{35r^9}+\frac{316}{35 M r^7}-\frac{48}{7 r^8}\right) P_4(\cos\theta )\bigg\}\ ,\\
   \phantom{a}\nonumber\\
g_{rr}=&-\frac{r}{r-2M}+\frac{2J^2}{r^2(r-2 M)} \left[\frac{1}{rM}\frac{(r-5 M)}{(r-2M)}P_2(\cos\theta )+1\right]+\frac{J^4}{(2M-r)^3}\left[\frac{152 M^2}{5 r^7}+\frac{9}{5 M^2 r^3}\right.\nonumber\\
&\left.-\frac{264 M}{5 r^6}-\frac{59}{5 M r^4}+\frac{196}{5 r^5}
+\left(-\frac{1464 M^2}{7 r^7}+\frac{52}{7 M^2 r^3}+\frac{1496 M}{7 r^6}-\frac{242}{7 M r^4}-\frac{106}{7 r^5}\right) P_2(\cos\theta )\nonumber\right.\\
&\left.+\left(\frac{2}{M^3 r^2}+\frac{2112 M^2}{35 r^7}-\frac{358}{35 M^2 r^3}-\frac{4512 M}{35 r^6}-\frac{8}{35 M r^4}+\frac{2616}{35 r^5}\right)P_4(\cos\theta )\right]\ ,\\
\phantom{a}\nonumber\\
g_{\theta\theta}=&-r^2+J^2 \left(\frac{2}{M r}+\frac{4}{r^2}\right) P_2(\cos\theta )-J^4 \left[\left(\frac{4}{7 M^2 r^4}+\frac{26}{7 M r^5}+\frac{68}{7 r^6}\right)P_2(\cos\theta )+\left(\frac{2}{M^3 r^3}\right.\right.\nonumber\\
&\left.\left.+\frac{162}{35 M^2 r^4}+\frac{24}{35 M r^5}+\frac{24}{35 r^6}\right) P_4(\cos\theta )\right]\ ,\\
\phantom{a}\nonumber\\
g_{\varphi\varphi}=&g_{\theta\theta}\sin\theta^2\ ,\\
\phantom{a}\nonumber\\
g_{t\varphi}=&\frac{2 J}{r}-J^3 \left[\frac{4}{5 M r^4}+\frac{12}{5 r^5}+\left(\frac{4}{Mr^4}+\frac{8}{r^5}\right)P_2(\cos\theta )+\left(\frac{2}{3 M^2 r^3}-\frac{2}{15 M r^4}-\frac{12}{5 r^5}\right) S_3(\theta)\right]\nonumber\\
&+J^5 \bigg\{\frac{24}{5 M r^8}+\frac{72}{5 r^9}-\frac{6}{7 M^3 r^6}-\frac{73}{35 M^2 r^7}-\left(\frac{2}{15 M^2 r^7}+\frac{2}{M r^8}+\frac{28}{5 r^9}\right) S_3(\theta )+\left[\frac{96}{35 M^2 r^7}\right.\nonumber\\
&\left.+\left(\frac{4}{3 M^3 r^6}+\frac{12}{5 M^2 r^7}-\frac{16}{3 M r^8}-\frac{48}{5 r^9}\right) S_3(\theta )+\frac{108}{7 Mr^8}+\frac{1016}{35 r^9}\right]P_2(\cos\theta )+\left(\frac{4}{M^3 r^6}\right.\nonumber\\
&\left.+\frac{324}{35 M^2 r^7}+\frac{48}{35 M r^8}+\frac{48}{35 r^9}\right) P_4(\cos\theta)+\frac{2}{5}\left(\frac{1}{M^4 r^5}+\frac{1}{7 M^3 r^6}-\frac{44}{7 M^2 r^7}+\frac{8}{ r^9}\right) S_5(\theta )\bigg\}\ .
\end{align}
\end{widetext}

\section{Coefficients of the small coupling }\label{app:coeff2}
In the following table, we show the numerical coefficients of various analytic expansions 
presented in the previous sections, as functions of the BH spin angular momentum and the EDGB coupling parameter. 
For the sake of clarity all the coefficients are rounded to the seventh digit. The exact expressions are available in a 
Supplemental Material {\scshape Mathematica} notebook. 

\begin{table*}
\begin{tabular}{c|ccccc}
\hline
\hline
&$r_\tn{h}/M$ & $r_\tn{ISCO}/M$ & ${\cal D}/M$ & 
$M\omega_{\varphi}\vert_\tn{ISCO}$ & $M\omega_{\theta}\vert_\tn{ISCO}$ \\
\hline
$a_0$ & 2.000000 & 6.000000 & 0 & 0.06804138 & 0.06804138 \\
$b_0 $& 0 & -3.265986 & 0 & 0.05092593 & 0.04166667 \\
$ c_0 $& -0.2500000 & -0.2962963 & 0 & 0.03717075 & 0.02488551 \\
$ d_0$ & 0 & -0.1436429 & 0 & 0.02797068 & 0.01521776 \\
$ e_0$ & -0.07812500 & -0.08957762 & 0 & 0.02176680 & 0.009504151 \\
 $f_0 $& 0 & -0.06362468 & 0 & 0.01744794 & 0.005997155 \\
$ a_1 $& 0 & 0 & 0.5000000 & 0 & 0 \\
$ b_1 $& 0 & 0 & 0 & 0 & 0 \\
$ c_1 $& 0 & 0 & -0.1250000 & 0 & 0 \\
$ d_1$ & 0 & 0 & 0 & 0 & 0 \\
$ e_1 $& 0 & 0 & -0.06250000 & 0 & 0 \\
$ f_1$ & 0 & 0 & 0 & 0 & 0 \\
$ a_2$ & -0.07656250 & -0.1047904 & 0.1520833 & 0.001563316 & 0.001563316 \\
$ b_2$ & 0 & -0.1201586 & 0 & 0.003733697 & 0.003267414 \\
$ c_2 $& -0.005273438 & 0.01442503 & -0.06562500 & 0.003913488 & 0.002948313 \\
$ d_2$ & 0 & 0.03108340 & 0 & 0.003160772 & 0.002041599 \\
$ e_2$ & 0.0007317631 & 0.02534504 & -0.01282676 & 0.002252579 & 0.001211371 \\
$ f_2 $& 0 & 0.03275054 & 0 & 0.001261845 & 0.0005010288 \\
 $a_3$ & -0.05482722 & -0.05057329 & 0.09658358 & 0.0007597788 & 0.0007597788 \\
$ b_3$ & 0 & -0.06564501 & 0 & 0.001933365 & 0.001708284 \\
 $c_3 $& 0.003374719 & 0.02695641 & -0.06503722 & 0.001759287 & 0.001360011 \\
$ d_3$ & 0 & 0.03707053 & 0 & 0.0009886223 & 0.0006954496 \\
 $e_3$ & -0.003268537 & 0.008889195 & 0.006444151 & 0.0005021134 & 0.0004182817 \\
 $f_3 $& 0 & 0.01125766 & 0 & 0.0001173107 & 0.0002713766 \\
$ a_4 $& -0.05139096 & -0.03780985 & 0.08178788 & 0.0005969829 & 0.0005969829 \\
$ b_4 $& 0 & -0.05431288 & 0 & 0.001640950 & 0.001457307 \\
 $c_4 $& 0.007351713 & 0.03876245 & -0.06717236 & 0.001340086 & 0.001049576 \\
$ d_4$ & 0 & 0.05336676 & 0 & 0.0003245650 & 0.0002579998 \\
$ e_4$ & -0.01308955 & -0.0002670172 & 0.02912691 & -0.00008270490 & 0.0001598391 \\
$ f_4 $& 0 & 0.00005147144 & 0 & -0.0002712206 & 0.0002346154 \\
 $a_5 $& -0.05266569 & -0.03321722 & 0.07886477 & 0.0005272127 & 0.0005272127 \\
$ b_5 $& 0 & -0.04890924 & 0 & 0.001475785 & 0.001313209 \\
$ c_5 $& 0.01223578 & 0.05011869 & -0.07508136 & 0.0009815575 & 0.0007719423 \\
$ d_5$ & 0 & 0.06775197 & 0 & -0.0003132155 & -0.0001796837 \\
$ e_5$ & -0.02668707 & -0.01513785 & 0.05662227 & -0.0005664351 & -0.00003872048 \\
$ f_5 $& 0 & -0.02079280 & 0 & -0.0003557189 & 0.0003967294 \\
$ a_6$ & -0.05753945 & -0.03250101 & 0.08245910 & 0.0005165825 & 0.0005165825 \\
$ b_6 $& 0 & -0.04824270 & 0 & 0.001458090 & 0.001298530 \\
$ c_6 $& 0.01812429 & 0.06363000 & -0.08788829 & 0.0007567590 & 0.0005928172 \\
$ d_6$ & 0 & 0.08594476 & 0 & -0.0008980528 & -0.0005936325 \\
$ e_6$ & -0.04690661 & -0.03465520 & 0.09290842 & -0.0009679387 & -0.0001794779 \\
$ f_6 $& 0 & -0.04866939 & 0 & -0.0002465314 & 0.0007018829 \\
$ a_7 $& -0.06565095 & -0.03416370 & 0.09098999 & 0.0005425110 & 0.0005425110 \\
$ b_7$ & 0 & -0.05069519 & 0 & 0.001533246 & 0.001365759 \\
$ c_7 $& 0.02592898 & 0.08041305 & -0.1064184 & 0.0005835871 & 0.0004504112 \\
$ d_7$ & 0 & 0.1086434 & 0 & -0.001525555 & -0.001044296 \\
$ e_7$ & -0.07619661 & -0.06177540 & 0.1427116 & -0.001328310 & -0.0002654697 \\
$ f_7$ & 0 & -0.08762124 & 0 & 0.0001137773 & 0.001223565 \\
\hline
\hline
\end{tabular}
\hspace{0.3cm}
\begin{tabular}{c|cc}
\hline
\hline
&$r_\tn{ergo}/M$ & $M^2 R_\tn{intr}$  \\
\hline
$ l_0 $& 2.000000 & 0.5000000 \\
$ m_0$ & 0 & 0 \\
 $n_0$ & -0.06640625 & -0.05859375 \\
$ p_0$ & -0.2500000 & -0.3750000 \\
$ q_0$ & -0.04687500 & -0.2343750 \\
$u_0$ & 0.03515625 & 0.1054688 \\
$l_1$ & 0 & 0 \\
$m_1$ & 0 & 0 \\
$n_1$& 0 & 0 \\
$p_1$ & 0 & 0 \\
$q_1$ & 0 & 0 \\
$u_1$ & 0 & 0 \\
$ l_2$ & -0.07656250 & 0.03828125 \\
$ m_2$ & 0.02239583 & -0.02182674 \\
$ n_2$ & -0.02390784 & -0.01729976 \\
$ p_2$ & -0.02766927 & -0.1164568 \\
$ q_2$ & 0.003249614 & -0.02236320 \\
$ u_2$ & 0.02138998 & 0.1302572 \\
 $l_3$ & -0.05482722 & 0.02741361 \\
 $m_3$ & 0.02660141 & -0.02038983 \\
 $n_3$ & -0.02726616 & 0.001655395 \\
 $p_3$ & -0.02322669 & -0.08694771 \\
 $q_3$ & 0.006734437 & 0.01982431 \\
 $u_3$ & 0.01726319 & 0.1306188 \\
 $l_4 $& -0.05139096 & 0.02789366 \\
 $m_4$ & 0.03391131 & -0.02622298 \\
 $n_4 $& -0.04523654 & 0.01648181 \\
 $p_4$ & -0.02655960 & -0.1007545 \\
 $q_4$ & 0.01024171 & 0.05758840 \\
 $u_4$ & 0.02190527 & 0.1906275 \\
 $l_5$ & -0.05266569 & 0.02948113 \\
 $m_5$ & 0.04318465 & -0.03204816 \\
 $n_5$ & -0.06780117 & 0.03771938 \\
 $p_5$ & -0.03094887 & -0.1122069 \\
 $q_5$ & 0.01536541 & 0.1033644 \\
 $u_5$ & 0.02574869 & 0.2495574 \\
 $l_6$ & -0.05753945 & 0.03296015 \\
 $m_6$ & 0.05574249 & -0.04017469 \\
 $n_6$ & -0.1009268 & 0.06678447 \\
 $p_6$ & -0.03761820 & -0.1310279 \\
 $q_6$ & 0.02256511 & 0.1634876 \\
$ u_6$ & 0.03145506 & 0.3323836 \\
 $l_7 $& -0.06565095 & 0.03820390 \\
 $m_7$ & 0.07278610 & -0.05103945 \\
 $n_7$ & -0.1481975 & 0.1073399 \\
 $p_7$ & -0.04685712 & -0.1565866 \\
 $q_7$ & 0.03308656 & 0.2435265 \\
 $u_7$ & 0.03891437 & 0.4418871 \\
\hline
\hline
\end{tabular}
\caption{ Numerical values of the 
coefficients of the expressions~(\ref{Eqrh}),(\ref{charge}),(\ref{Isco}),
 (\ref{omegaphi})-(\ref{omegatheta}) (left panel), and 
of the  ergosphere and intrinsic curvature radius~(\ref{EqERGO})], 
(\ref{curvrad}) (right panel).}
\label{TableCoeff}
\end{table*}

\bibliography{bibnote}

\end{document}